\documentstyle[aps,prl,epsf]{revtex}
\twocolumn

\begin{document}
\draft
\twocolumn[\hsize\textwidth\columnwidth\hsize\csname @twocolumnfalse\endcsname

\title{Evolution of spectral function in a doped Mott insulator :
surface vs. bulk contributions}

\author{K. Maiti, Priya Mahadevan\cite{phy}, and D.D. Sarma\cite{JNC}}
\address{Solid State and Structural Chemistry Unit, Indian 
Institute of Science, Bangalore 560 012, India}

\date{\today}

\maketitle

\begin{abstract}
We study the evolution of the spectral function with progressive
hole doping in a Mott insulator, La$_{1-x}$Ca$_x$VO$_3$ with $x$ =
0.0 - 0.5. The spectral features indicate a bulk-to-surface
metal-insulator transition in this system. Doping dependent changes
in the bulk electronic structure are shown to be incompatible with
existing theoretical predictions. An empirical description based on 
the single parameter, $U/W$, is shown to describe consistently the
spectral evolution.
\end{abstract}
]

It is well understood that the insulating ground state of transition
metal compounds with integral occupancies of 3$d$ levels (Mott
insulators) arise from strong electron correlation effects.  Recent
years have seen a phenomenal resurgence of interest in studying the
properties of such systems doped with charge carriers representing
fractional 3$d$ occupancies, leading to the discovery of many exotic
properties such as the high temperature superconductivity
\cite{hitcsc} and colossal magnetoresistance\cite{cmr}. A
considerable amount of effort has been directed to modelling these
systems theoretically, which is crucial for understanding such
properties. In recent times, there are very specific predictions
about the evolution of the spectral functions in doped Mott
insulators based on the Hubbard model in the limit of infinite
dimension\cite{infd1,infd2}. Considerably different ansatz, such as
slave boson\cite{slvbn}, dynamical mean field theories\cite{dft} and
exact calculations\cite{ec} for finite systems, yield qualitatively
similar results. These predictions are most useful, since the
spectral function can be directly measured experimentally by
photoemission spectroscopy and thus, provide a very convenient
testing ground for the suitability of the model. In order to
investigate the spectral evolution as a function of doping we have
studied La$_{1-x}$Ca$_x$VO$_3$ for $x$ = 0.0 - 0.5. LaVO$_3$ is a
Mott insulator with the $d^2$ electronic configuration.  Ca
substitution dopes holes into the system continuously varying the
electron count from 2 to 1.5 per V ion for the compositions studied.
Transport and magnetic measurements\cite{resmag} have shown that the
system is an antiferromagnetic insulator for $x$ $<$ 0.2, but is
Pauli paramagnetic metal for $x \ge$ 0.2. Moreover, there is no
significant structural modification suggesting nearly constant
bandwidth ($W$), with a little ($\le$ 10\%) change in Hubbard $U$ across the
compositions studied \cite{bocquet}.  Hence, the change in the
electronic structure in this system is primarily driven by the
changes in the doping level and thus, provide a suitable case to
investigate the spectral evolution in a doped Mott insulator. We
observe the surface electronic structure in this system to be
considerably different from the bulk one and provide an algorithm to
extract the true bulk electronic structure from the total spectra
using photon energy dependence of the spectral functions. The
extracted bulk spectra show that existing theories based on the
Hubbard model are insufficient to describe the experimentally
observed spectral functions for various dopings. We provide an
empirical approach to describe the observed evolution of the spectral
function and discuss the implications of our findings.

Polycrystalline samples of La$_{1-x}$Ca$_x$VO$_3$ were prepared as
reported elsewhere\cite{resmag}, from congruently molten states
giving rise to large grains with strong intergrain bonding and about
1\% of impurity phases in the grain boundaries.  The x-ray
photoemission (XP) and ultraviolet photoemission (UP) spectroscopic
measurements were carried out at liquid nitrogen temperature in a
vacuum of 4$\times$10$^{-10}$ mbar with resolutions of 0.8 eV and
0.08 eV respectively. The cleanliness of the sample surface was
maintained by periodical scraping with an alumina file {\it in situ}
and was monitored by O 1$s$ and C 1$s$ spectral regions in XP and the
characteristic region (9 - 12 eV binding energy) in UP measurements.
Reproducibility of the spectra with repeated scrapings was confirmed
for each composition. We believe that the spectra are representative
of angle integrated spectra, since they are obtained over a wide
($\pm$ 10$^o$) acceptance angle and were always independent of sample
rotation about the surface normal. The bulk spectra were primarily
extracted from the XP spectra, which by their nature average over a
large part of the Brillouin zone. We implicitly assume that the
incident photon energy-dependent cross-sections and the combined
resolution do not seriously change the spectra over our narrow (3 eV)
binding energy window; we substantiated this assumption explicitly
for LaVO$_3$. The derived spectra for the bulk, within our resolution
and over the relatively large energy range investigated here, are not
affected by the disorder due to scraping. The bulk and surface compositions
were found to be in agreement with nominal compositions from energy 
dispersive $x$-ray analysis and the intensity ratio of the core level
spectra of the componenets, respectively.

In Fig.1 we show the He {\scriptsize I} UP spectra for all the
compounds near the V 3$d$ emission region. Each spectrum is dominated
by a feature centered at about 1.5 eV below $E_F$, with no intensity
at the Fermi level ($E_F$) for $x$ = 0.0 and 0.1. This feature is
normally termed as incoherent feature \cite{incoherent} being the
spectral signature of the lower Hubbard band (LHB) and corresponds to
electron states essentially localized due to electron correlations.
The spectral feature growing at $E_F$ (coherent peak) with increasing
$x$ ($x \ge 0.2$) represents the delocalized conduction electrons.
Thus, these observations show a metal-insulator (MI) transition at
$x$ = 0.2 in agreement with the transport properties\cite{resmag}.
However, the complete dominance of the incoherent feature suggests an
overwhelming presence of correlation effects nearly localizing charge
carriers even in the $x = 0.5$ sample in contrast to the observed
physical properties\cite{resmag}.  As we shall show now, at least a
part of this discrepancy arises from a surface transition of the
electronic structure in these systems.

The surface sensitivity of these electronic spectra can be varied by
changing the exciting photon energy, since the escape depth
($\lambda$) of the electrons depends sensitively on their kinetic
energy\cite{mfp}. We use this fact to delineate the surface and bulk
electronic structures by recording the spectra using He~{\scriptsize
I} (21.2 eV), He~{\scriptsize II} (40.8 eV) and monochromatized Al
K$\alpha$ (1486.6 eV) sources; among these the spectrum excited with
He {\scriptsize II} radiation is expected to be the most surface
sensitive, while the Al K$\alpha$ spectrum the most bulk sensitive.
We show the spectra for each of the metallic compositions in Fig.2,
after subtracting the O $p$ contributions (solid line in the inset of
Fig.1) appearing at binding energies higher than 3.5 eV following the
procedures in Refs.\cite{morikawa,fujimori}.  For each sample, the
spectrum recorded with He~{\scriptsize II} radiation exhibits the
single incoherent spectral signature with no intensity at $E_F$
indicating absence of any coherent state. This clearly demonstrates
that the electronic structure in the surface region of these samples
remains localized. This most probably arises from an enhanced
correlation effect near the surface region due to the reduced
dimensionality at the surface and/or subtle changes of surface
geometry (surface reconstruction) and consequently a reduced
bandwidth, even for $x = 0.5$ sample. This is not entirely surprising
since the critical concentration even for the bulk MI transition in
closely related, but slightly distorted, series Y$_{1-x}$Ca$_x$VO$_3$
is indeed $x= 0.5$ \cite{ycavo3}. The finite spectral weight in the
coherent feature in the He~{\scriptsize I} spectrum increases further
in the Al K$\alpha$ excited signal for each of the samples,
consistent with an increasing bulk sensitivity of the technique. This
substantiates further the view that the bulk electronic structure
represents a metallic state at these compositions, though the surface
layer has only localized states. Hence, to probe the true bulk
spectral function, it is however necessary to separate the surface
contribution from the total spectra in Fig.2.

The total normalized spectral intensity F($\epsilon$) from the sample
at normal emission is given by $$F(\epsilon) = F^s(\epsilon)
(1-e^{-d/\lambda}) + F^b(\epsilon) e^{-d/\lambda}$$ where,
$F^s(\epsilon)$ (given by the He~{\scriptsize II} spectra) and
$F^b(\epsilon)$ are the normalized surface and bulk spectral
functions respectively and $d$ is the thickness of the insulating
surface layer. Expressing $e^{d/\lambda}$ as $\alpha$, we readily
find that $F^b(\epsilon)$ is given by $[\alpha.F(\epsilon) -
(\alpha-1).F^s(\epsilon)]$.  If $\alpha$, which is a function of the
photon energy via $\lambda$, would be known, it would be straight
forward to obtain the bulk spectrum, $F^b(\epsilon)$, from the above
equation. In absence of a prior knowledge of $\alpha$ however, we
note that the spectral shape of $F^b(\epsilon)$ representing the bulk
electronic structure is independent of the photon energy. Thus, the
extracted $F^b(\epsilon)$ from the Al K$\alpha$ spectra must be
consistent with that extracted from the He~{\scriptsize I} spectrum.
This consistency requirement can be cast into a least-squared-error
procedure to estimate the values of $\alpha$ corresponding to He
{\scriptsize I} and Al K$\alpha$ photon energies.  Thus obtained
values of $\alpha_{He I}$ = 2.95 and $\alpha_{XPS}$ = 1.38 correspond
to $d/\lambda_{XPS}$ = 0.32 and $d/\lambda_{He I}$ = 1.08 leading to
a reasonable $\lambda_{XPS}/\lambda_{He I}$ $\approx$ 3.4.  Moreover,
using the estimates of $\lambda$ from the published
literature\cite{mfp,penedpt1,penedpt2}, the present result suggests
that the thickness, $d$, of the insulating surface layer is
approximately 8 \AA,\ which is roughly the dimension of the unit
cell.

In inset I of Fig.3 we demonstrate the subtraction procedure for
obtaining the bulk spectra\cite{proc}. Thus obtained bulk electronic
structures for the metallic compositions are shown in Fig.3 by open
circles.  The spectral weights show a systematic increase at $E_F$
with increasing Ca, consistent with increasing conductivity in
the series. However, the V 3$d$ emission is seen to be spread to
about 3 eV below $E_F$ with the highest intensity in the vicinity of
1.5 eV. Our band structure calculations for both LaVO$_3$ and
CaVO$_3$ however suggest that the electronic density of
states in the occupied part has a peak at $E_F$ with a total spread
of about 1 eV. Thus, it is clear that the bulk electronic structures
as characterized by the spectra in Fig.3 are strongly influenced by
the electron correlation effect, leading to substantial incoherent
spectral signatures giving rise to the peak near 1.5 eV and an
extension of the emission down to 3 eV below $E_F$. The
doping-dependent spectral function has been calculated within the
single band Hubbard model both by dynamical mean field approximations
\cite{dft} and exact diagonalization techniques\cite{ec}.  These
results indeed suggest the presence of the incoherent and coherent
features in the spectra.  However, any reasonable value of $U/W$ with
finite doping shows a much weaker  contribution  of the incoherent
feature as compared to our experimental results. This weakened effect
of correlation in the presence of large doping within a single band
Hubbard model is understandable, since such doping reduces the
average number of electrons per site considerably from 1.0. In the
present case, there are 1.5 electrons per V site even for the highest
doped case (La$_{0.5}$Ca$_{0.5}$VO$_3$). In order to understand the
effects of correlation in such a multiband case, we have calculated
the spectral function for a degenerate Hubbard model \cite{dhbref},
since such results are not available in the literature. We show the
resolution broadened spectral functions for various values of $U/W$
in inset II of Fig.3 exhibiting a complete dominance of the coherent
feature and the presence of only weak incoherent features above 1.0
eV. Thus, the inability of the Hubbard model to account for the
experimentally observed intense incoherent spectral features in these
doped Mott insulators\cite{dft,ec} cannot be attributed to the
multiband nature of the problem.

In this context, it is important to realize that the end members of
La$_{1-x}$Ca$_x$VO$_3$ have fundamentally different electronic
properties, one (LaVO$_3$) being a Mott insulator, while the other
(CaVO$_3$) is a correlated metal. Moreover, we note that the peak
position near 1.5 eV and its extension down to 3 eV is remarkably
similar to the spectral signature of LaVO$_3$ (see Fig.3). The
spectral weight shifted nearer to $E_F$ in the metallic samples
compared to LaVO$_3$ is clearly the signature of the coherent feature
growing in relative intensity with progressive doping of the Mott
insulator. It turns out that we can model this coherent part of the
spectral signature with {\it ab initio} band structure calculations
for CaVO$_3$ corrected for correlation effects within the
perturbative treatment of the self-energy \cite{treglia}. 
Considering linear contributions of the spectral features arising
from LaVO$_3$ and CaVO$_3$, the general sum rule for the
conservation of electron number completely determines the relative
contributions.  This can be expressed as the total measured spectrum
$\rho(\epsilon) = 2(1-x)\rho_{La}(\epsilon) + x\rho_{Ca}(\epsilon)$.
Here, $\rho_{La}(\epsilon)$ is the normalized spectral function
of the LaVO$_3$ part and is completely determined from the
experimental XP spectrum of LaVO$_3$. $x$ is also fixed by the
composition of the compound La$_{1-x}$Ca$_x$VO$_3$ and ensures a
total electron count per V ion equal to (2-$x$).
$\rho_{Ca}(\epsilon)$ is calculated from the 
band structure results corrected up to second order in $U/W$.
Thus, $\rho(\epsilon)$ depends on a {\it single} parameter, $U/W$,
with all other contributions being {\it a-priori} fixed.  Noting the
possible limitations of any parametric fitting procedures, we fix
even the value of this parameter ($U/W$ = 0.5) by requiring that the
band structure results for CaVO$_3$ when corrected for
correlation effects via self energy calculation, yields the correct
$m^\star/m_b$ = 2.1 \cite{inoue}. Thus, the calculated total spectrum
$\rho(\epsilon)$ is completely devoid of any adjustable parameter.
The change in the spectral features is determined
solely  by the value of $x$. The resulting $\rho(\epsilon)$ are
shown in Fig.3 by the solid lines for each of the compositions
establishing a remarkable agreement between the experiment and the
parameter-free simulation. This result suggests that LaVO$_3$ and
CaVO$_3$ somehow retain their characteristic electronic structures
even for the intermediate compositions indicating a dominance of
local interactions in determining the electronic structures; this is
however a reasonable expectation for strongly correlated systems.

In conclusion, it is evident from our results that the surface
electronic structure of early transition metal oxides can be
$qualitatively$ different from the bulk one. This realization is
essential in order to critically discuss and evaluate the
experimental electronic structure in terms of the existing many-body
theories and various bulk sensitive low-energy properties. We present
an algorithm to extract the bulk related spectra from the total
spectra using photon energy dependent measurements. It is shown that
all known theoretical results based on the single band Hubbard model
\cite{dft,ec} as well as the present multiband Hubbard model, are
clearly insufficient to explain the experimentally observed bulk
spectral functions for $any$ value of $U/W$. It is then remarkable
and significant that a simple additive description of the spectral
functions of the end members, LaVO$_3$ and CaVO$_3$, though empirical
and speculative, is qualitatively and quantitatively successful for
all the compositions without using any adjustable parameter and
therefore deserves serious considerations in terms of rigorous
microscopic theories for such systems. A direct consequence of this
observation would be that the electronic structure of such a system
is intrinsically inhomogeneous and cannot be described within a
homogeneous model, such as the Hubbard model. This is in contrast to
the usual practice in the recent times to interpret the properties of
doped Mott insulators in terms of the Hubbard model. While the
presented empirical approach appears to be consistent with the
spectral evolution as a function of doping of the Mott insulator
including both the coherent and the incoherent parts, it remains to
be seen if in future a proper theoretical model would provide a
rigorous basis for understanding the physical properties of these
systems. 

We thank G. Kotliar and P. Mazumdar for useful discussions and C.N.R.
Rao for continued support. We thank the Dept. of Science and 
Technology and the Council of Scientific
and Industrial Research, Govt. of India for financial assistance.

\section{figure captions}
Fig.1 He {\scriptsize I} spectra of 
La$_{1-x}$Ca$_x$VO$_3$. Expanded ($\times$3) view of near $E_F$
region is also shown. Inset shows expected tailing of O $p$ states.

Fig.2 Normalized (equal area) He {\scriptsize I}, He {\scriptsize II}
and XP valence band spectra of La$_{1-x}$Ca$_x$VO$_3$ for $x$ = 0.3,
0.4 and 0.5.

Fig.3 Bulk spectra (open circles) of La$_{1-x}$Ca$_x$VO$_3$.
The solid lines show the synthesized spectra.
The corresponding spectra of LaVO$_3$(solid
circles) and CaVO$_3$(dashed lines) are also shown. Inset I shows the
surface and bulk components of the spectral function for He
{\scriptsize I} and XP spectra for $x$ = 0.5(see \cite{proc}). Inset
II shows the results obtained from multiband Hubbard model for
various values of $U/W$.


\begin{thebibliography}{99}
\bibitem{proc} The raw UP spectra are convoluted with Gaussians to
account for different energy resolutions; normalized He {\scriptsize
I} and Al K$\alpha$ spectra corresponding to $F(\epsilon)$ are
represented by cross(+) in the inset I of Fig.3 while the He
{\scriptsize II} spectra multiplied by ($\alpha$-1) (the surface
component) is shown by dot-dashed line. The resulting bulk component
of the spectra is obtained from the XP spectrum by subtraction, shown
as open circles. The same bulk spectrum also reproduces (solid line)
the experimental He {\scriptsize I} spectrum well (the consistency
requirement).
\bibitem{dhbref} The model consists of 4 sites with all five
$d$-orbitals and 1.5 electrons per site. The hoppings are given by
Slater-Koster parametrizations with $(dd\sigma) : (dd\pi) :
(dd\delta) = -2.0 : 1.0 : -0.2$ with a total bandwidth $W$ = 2.4 eV.
\end{thebibliography}
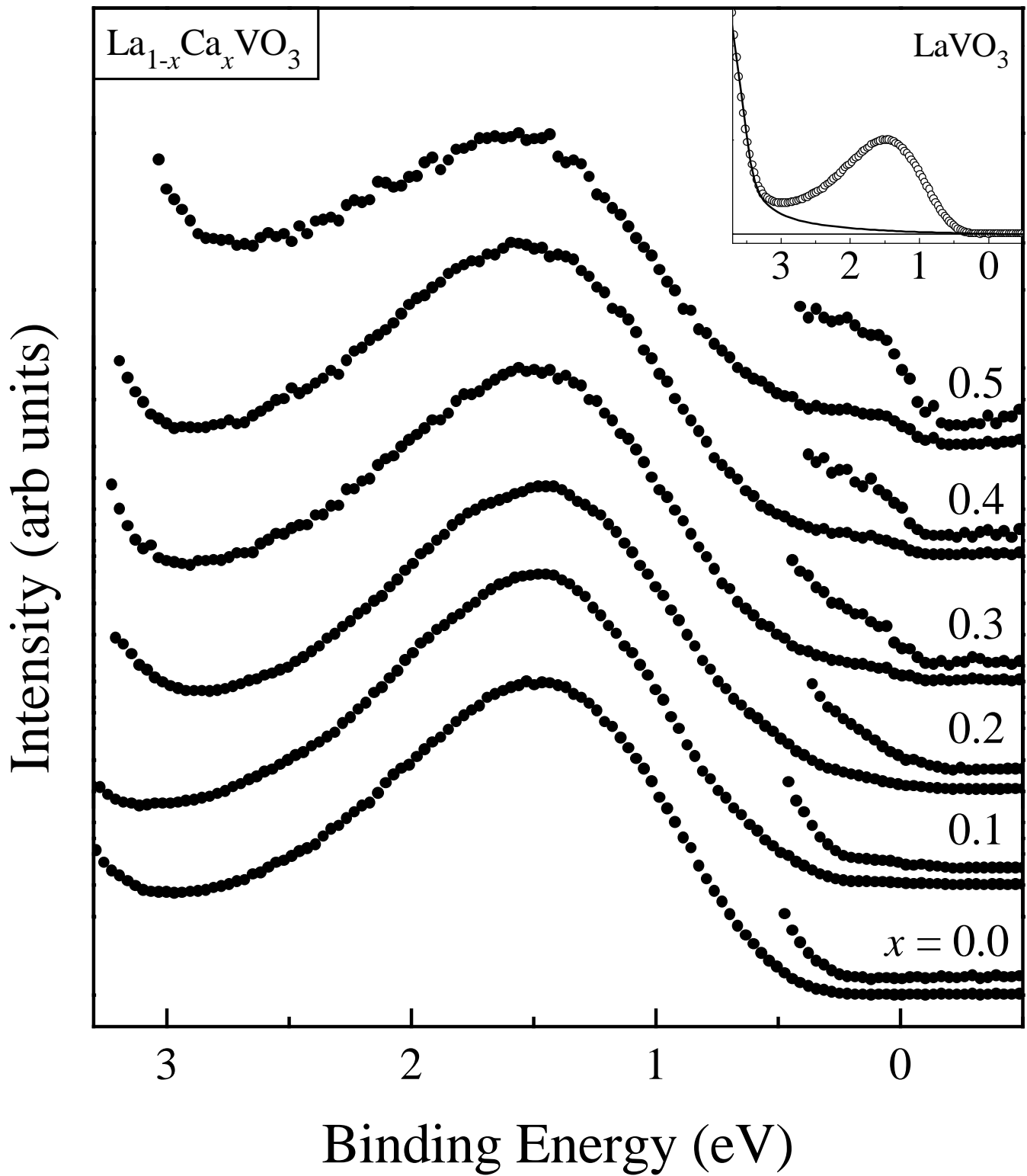
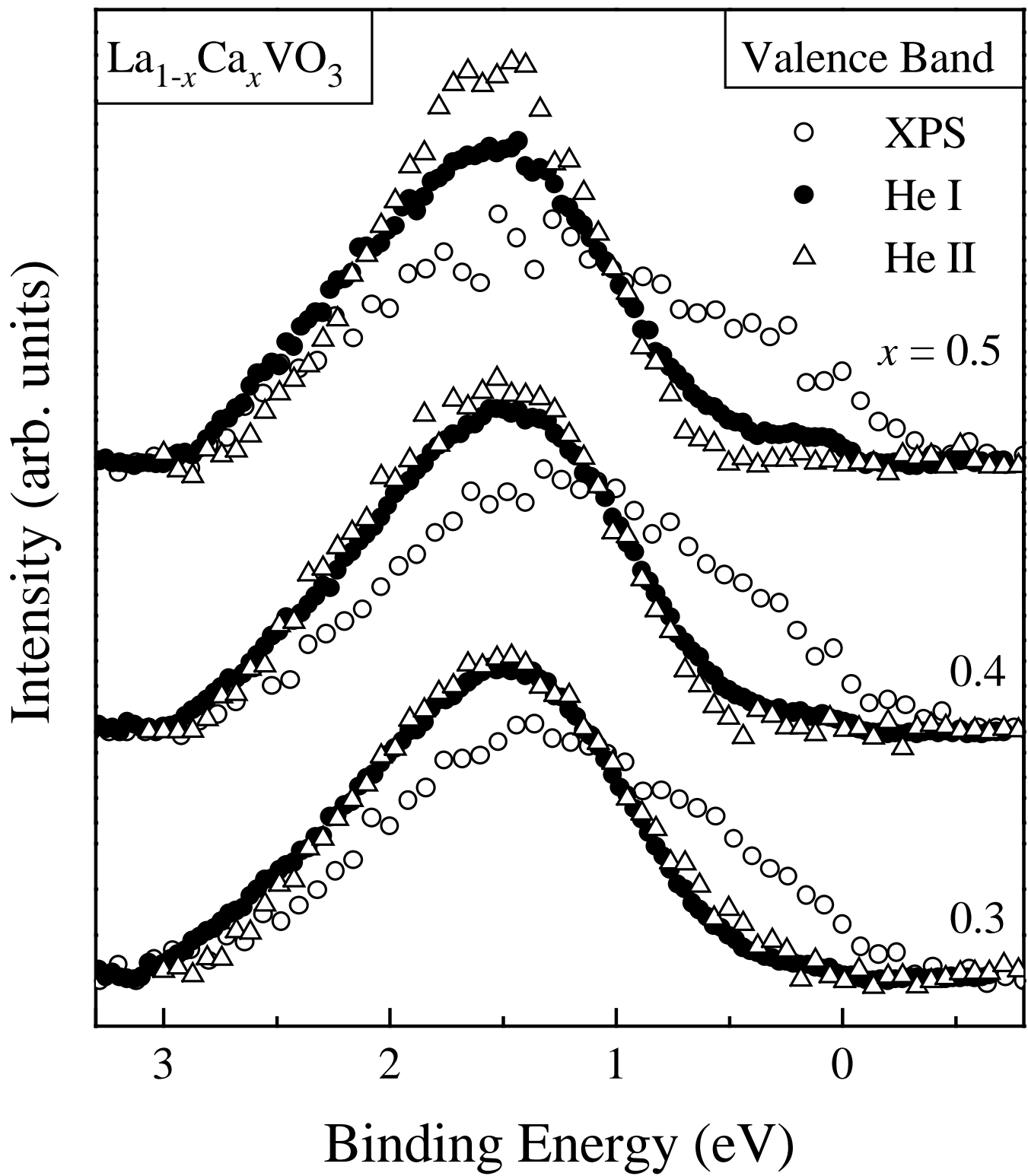
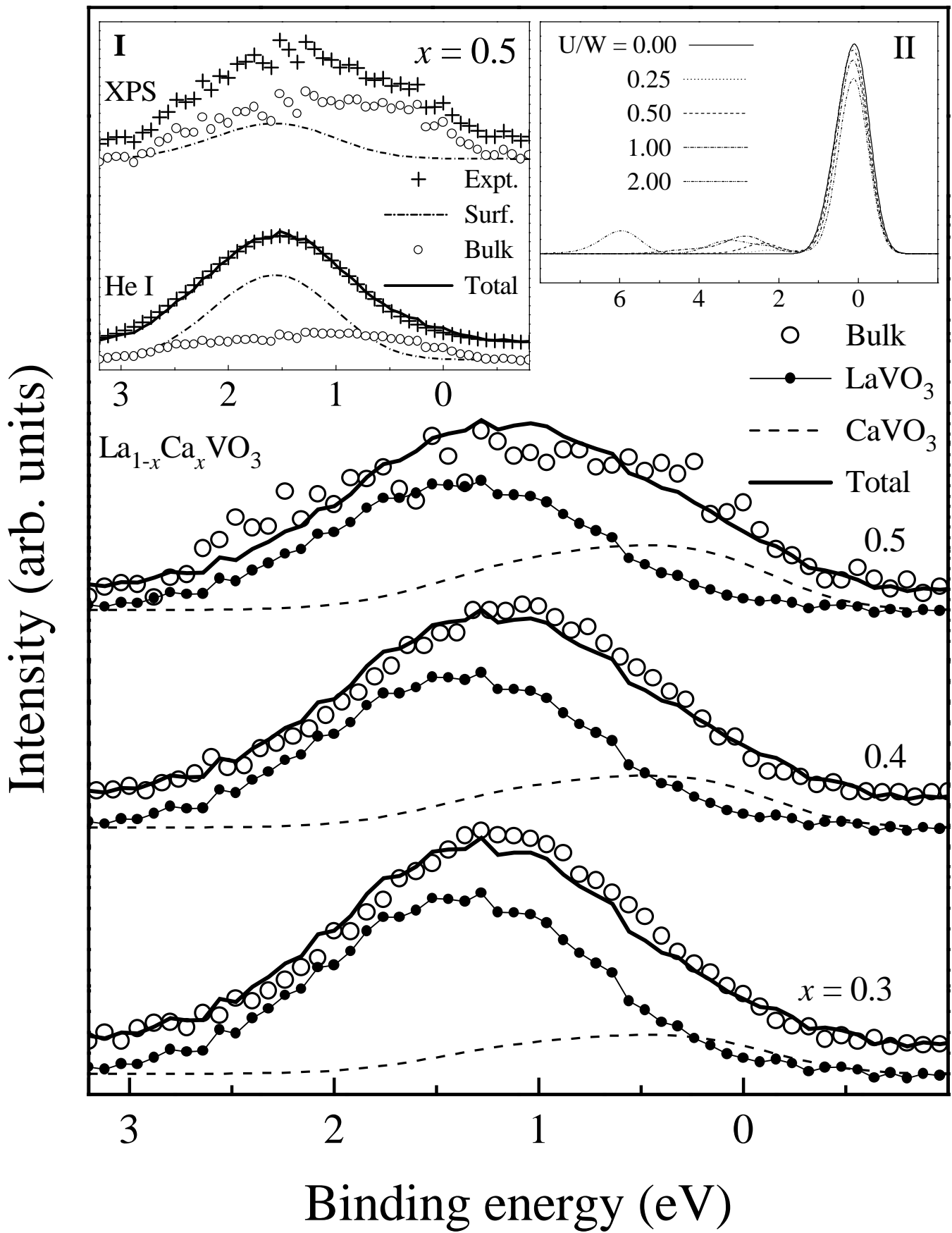
\end{document}